\documentclass[aps,twocolumn,preprintnumbers,amsmath,amssymb,amsfonts,groupedaddress,superscriptaddress,showpacs,showkeys,floatfix]{revtex4-1}
\usepackage{amsmath}
\usepackage{amssymb}
\usepackage{amsfonts}
\usepackage{algorithm}
\usepackage{algorithmicx}
\usepackage{algpseudocode}
\usepackage{graphicx}
\usepackage{threeparttable}
\usepackage{multirow}
\usepackage{array}
\usepackage{xcolor}
\usepackage[colorlinks,
            linkcolor=blue,
            anchorcolor=blue,
            citecolor=blue,
            urlcolor=blue
            ]{hyperref}

\newcommand{\bmath}{\begin{mathletters}}
\newcommand{\emath}{\end{mathletters}}
\newcommand{\be}{\begin{eqnarray}}
\newcommand{\ee}{\end{eqnarray}}
\newcommand{\ba}{\begin{array}}
\newcommand{\ea}{\end{array}}

\newcommand{\no}{\nonumber}

\definecolor{blue}{rgb}{0,0,1}
\definecolor{red}{rgb}{1,0,0}
\definecolor{green}{rgb}{0,1,0}
\definecolor{black}{rgb}{0,0,0}

\begin{document}
\title{Determination of Molecular Energies via Quantum Imaginary Time Evolution \\  in a Superconducting Qubit System}

\author{Zhiwen Zong}
\thanks{These authors have contributed equally to this work.}
\affiliation{Zhejiang Province Key Laboratory of Quantum Technology and Device, School of Physics, Zhejiang University, Hangzhou, 310027, China}

\author{Sainan Huai}
\thanks{These authors have contributed equally to this work.}
\affiliation{Tencent Quantum Laboratory, Tencent, Shenzhen, Guangdong 518057, China}

\author{Tianqi Cai}
\thanks{These authors have contributed equally to this work.}
\affiliation{Tencent Quantum Laboratory, Tencent, Shenzhen, Guangdong 518057, China}

\author{Wenyan Jin}
\affiliation{Zhejiang Province Key Laboratory of Quantum Technology and Device, School of Physics, Zhejiang University, Hangzhou, 310027, China}

\author{Ze Zhan}
\affiliation{Zhejiang Province Key Laboratory of Quantum Technology and Device, School of Physics, Zhejiang University, Hangzhou, 310027, China}

\author{Zhenxing Zhang}
\affiliation{Tencent Quantum Laboratory, Tencent, Shenzhen, Guangdong 518057, China}

\author{Kunliang Bu}
\affiliation{Tencent Quantum Laboratory, Tencent, Shenzhen, Guangdong 518057, China}

\author{Liyang Sui}
\affiliation{Zhejiang Province Key Laboratory of Quantum Technology and Device, School of Physics, Zhejiang University, Hangzhou, 310027, China}

\author{Ying Fei}
\affiliation{Zhejiang Province Key Laboratory of Quantum Technology and Device, School of Physics, Zhejiang University, Hangzhou, 310027, China}

\author{Yicong Zheng}\email{yicongzheng@tencent.com}
\affiliation{Tencent Quantum Laboratory, Tencent, Shenzhen, Guangdong 518057, China}

\author{Shengyu Zhang}
\affiliation{Tencent Quantum Laboratory, Tencent, Shenzhen, Guangdong 518057, China}

\author{Jianlan Wu}\email{jianlanwu@zju.edu.cn}
\affiliation{Zhejiang Province Key Laboratory of Quantum Technology and Device, School of Physics, Zhejiang University, Hangzhou, 310027, China}

\author{Yi Yin}\email{yiyin@zju.edu.cn}
\affiliation{Zhejiang Province Key Laboratory of Quantum Technology and Device, School of Physics, Zhejiang University, Hangzhou, 310027, China}

\begin{abstract}
As a valid tool for solving ground state problems, imaginary time evolution (ITE) is widely used in physical and chemical simulations. Different ITE-based algorithms in their quantum counterpart have recently been proposed and applied to some real systems. We experimentally realize the variational-based quantum imaginary time evolution (QITE) algorithm to simulate the ground state energy of hydrogen ($\mathrm{H_2}$) and lithium hydride (LiH) molecules in a superconducting qubit system. The H$_2$ molecule is directly simulated using the 3-qubit circuit with unitary-coupled clusters (UCC) ansatz. We also combine QITE with the cluster mean-field (CMF) method to obtain an effective Hamiltonian. The LiH molecule is correspondingly simulated using the 3-qubit circuit with hardware-efficient ansatz. For comparison, the LiH molecule is also directly simulated using the 4-qubit circuit with UCC ansatz at the equilibrium point. All the experimental results show a convergence within 4 iterations, with high-fidelity ground state energy obtained.  For a more complex system in the future, the CMF may allow further grouping of interactions to obtain an effective Hamiltonian, then the hybrid QITE algorithm can possibly simulate a relatively large-scale system with fewer qubits.
\end{abstract}

\maketitle

\section{Introduction}\label{sec1}
Calculating the ground state energy of a target Hamiltonian is one important application of a universal quantum computer~\cite{AbramsPRL97}. Some physical and chemical models have been preliminarily studied in noisy intermediate-scale quantum (NISQ) devices~\cite{Preskill}. The essence of those models lies in the interaction between the atomic nuclei and electrons, equivalent to solving the Schr$\ddot{\mathrm{o}}$dinger equation of a many-body system. Due to the difficulty of obtaining an exact analytical solution, researchers have introduced a variety of approximations. For example, the Hartree-Fock and post-Hartree-Fock methods can effectively handle isolated systems with weak and strong electron interactions, respectively~\cite{Townsend19}.  Furthermore, relevant quantum algorithms should be designed to process different theoretical models on a practical quantum computer. Variational quantum eigensolver (VQE)~\cite{PeruzzoNc14, HarperNc19, AbhinavNat17, RubinScience20}, quantum phase estimation~\cite{KitaevArxiv, AspuruScience05}, quantum Monte Carlo~\cite{Acioli97, WilliamNat22}, and quantum adiabatic optimization~\cite{FarhiScience01, Lucas14} are typical algorithms that have been proposed and applied to solve the ground state energy of molecular systems. Because of its simplicity and scalability, VQE is one of the most widely used algorithms, whose core idea is to construct a parameterized circuit and find the minimum of a target Hamiltonian through variational iterations. The VQE algorithm has been successfully implemented in many experiments, including IBM's simulation of BeH$_2$ with six qubits, Google's simulation of H$_{12}$ with twelve qubits~\cite{AbhinavNat17, RubinScience20} and others.

Despite its many advantages, the VQE algorithm mainly relies on mathematical gradient descent optimization, which may sometimes be difficult to provide an intuitive and instructive optimization path. Here we discuss a variational-based quantum imaginary time evolution (QITE) method~\cite {McArdleNPJQ19}, with the optimization path guided by physical principles. The imaginary time evolution (ITE) converts the real-time in the Schr$\ddot{\mathrm{o}}$dinger equation to an imaginary time through a Wick rotation~\cite{Wick1954}. For any initial wave function that is not orthogonal to the ground state, the high-energy components are rapidly filtered during the ITE, and only the ground state information is retained. However, for a non-commuting local term $h[m]$ of the system, the non-unitary ITE operator $e^{-h[m]\Delta \tau}$ makes it impossible for ITE to be directly applied to quantum computers. One popular idea is to find a Hermitian operator $\zeta[m]$ in a domain space to make $e^{-i\zeta[m]\Delta \tau} \rightarrow e^{-h[m]\Delta \tau}$~\cite{MottaNp20, AydenizNPJQ20, HirshPRXQ22}. Another method is to apply the variational principle directly to the ITE equation, such that the state evolution can be converted into an update of variational gate parameters. With McLachlan's variational principle, the gate parameters can all be updated with real numbers, making it especially applicable in a quantum circuit~\cite{McLachlan64, XiaoYuan19, YxyPRXQ21}. Recently, researchers also apply McLachlan's variational principle to a double-exponential function and introduce a QITE-like quantum iterative power algorithm (QIPA)~\cite{KyawArxiv22}. These QITE circuits can be solved by introducing an ancillary qubit~\cite{ArturPRL02, YingLiPRX17}, which may partially avoid crosstalk-induced readout errors in a multi-qubit system.

In this work, we focus on experimentally implementing the variational-based QITE in a superconducting multi-qubit chip. Two typical circuits, the unitary-coupled-clusters (UCC)~\cite{Andrew06, JonathanQST19} and hardware-efficient (HE) ansatzes~\cite{AbhinavNat17}, are selected for the QITE algorithm. We choose three or four qubits from the chip to determine the ground state energy of two sample molecules, hydrogen ($\mathrm{H_2}$) and lithium hydride (LiH). For $\mathrm{H_2}$, we directly apply the QITE method to simulate the total Hamiltonian. For the relatively complex LiH, we combine the QITE together with the cluster-mean-field (CMF) method to reduce the consumption of qubit resources~\cite {ZhanPRR22}. The 3-qubit experiment with hardware-efficient ansatz is implemented to solve the CMF-reduced Hamiltonian of LiH. A comparative 4-qubit experiment with UCC ansatz is also implemented to simulate the original Hamiltonian of LiH. We experimentally verify that the ground state energy of both molecules can quickly converge within only 4 iterations with the variational-based QITE circuits.

\section{Method of QITE}\label{sec2}
For a given system with time-independent Hamiltonian $H$, the state evolution follows the stationary Schr$\ddot{\mathrm{o}}$dinger equation, $|\psi(t)\rangle = e^{-iHt}|\psi(0)\rangle \approx \prod e^{-iHdt}|\psi(0)\rangle$. By artificially replacing $t$ with $i\tau$, we can transform the state to one with imaginary time evolution as
\begin{eqnarray}
|\psi(\tau)\rangle = c(\tau)e^{-H\tau}|\psi(0)\rangle,
\label{eqs_001}
\end{eqnarray}
in which $\tau$ is an imaginary time and $c(\tau) = \langle\psi(0)|e^{-2H\tau}|\psi(0)\rangle^{-1/2}$ is the normalization coefficient. Eigenstates and eigenenergies of the system are assumed to be non-degenerate $\{|\phi_n\rangle\}$ and $\{E_n\}$ ($n=0,1,\cdots,N-1;E_0<E_1<\cdots<E_{N-1}$). Substituting an arbitrary initial state $|\psi(0)\rangle=\sum_{i}\alpha_i|\phi_i\rangle$ into Eq.~(\ref{eqs_001}), we can obtain the state under eigenbases
\begin{eqnarray}
|\psi(\tau)\rangle=(1+\sum_{i=1}^{N-1}\frac{|\alpha_i|^2}{|\alpha_0|^2}e^{-2(E_i-E_0)\tau})^{-1/2}\no \\
(\frac{\alpha_0}{|\alpha_0|}|\phi_0\rangle+\sum_{i=1}^{N-1}\frac{\alpha_i}{|\alpha_0|}e^{-(E_i-E_0)\tau}|\phi_i\rangle).
\label{eqs_002}
\end{eqnarray}
This equation shows that as long as the initial wave function $|\psi(0)\rangle$ is not orthogonal to the ground state $|\phi_0\rangle$, namely $\alpha_0 \neq 0$, whatever wave function we choose will finally evolve to the ground state, $\lim_{\tau \to \infty} |\psi(\tau)\rangle = |\phi_0\rangle$. According to Eq.~(\ref{eqs_001}), we can also obtain the Schr$\ddot{\mathrm{o}}$dinger equation in the Wick rotation form
\begin{eqnarray}
\frac{\partial |\psi(\tau)\rangle }{\partial \tau}=-(H-E(\tau))|\psi(\tau)\rangle,
\label{eqs_003}
\end{eqnarray}
with $E(\tau) = \frac{\partial c(\tau)}{\partial \tau}c(\tau)^{-1}=\langle\psi(\tau)|H|\psi(\tau)\rangle$ denoting the average system energy at time $\tau$.

The actual quantum state at $\tau$ can be approximated by parameterization, $|\psi(\tau)\rangle \rightarrow |\psi({\boldsymbol\theta(\tau)})\rangle=|\psi(\theta_1,\theta_2,\cdots,\theta_\gamma)\rangle$. Applying McLachlan's variational principle to Eq.~(\ref{eqs_003}), $\delta||(\frac{\partial }{\partial \tau}+(H-E(\tau))|\psi(\tau)\rangle||=0$, we can transform the time evolution into the ansatz space with resulting linear equations $\sum_{j}A_{ij} \dot{\boldsymbol\theta_j} =B_i$, in which $A_{ij}=\mathrm{Re}[\frac{\partial \langle\psi(\boldsymbol\theta)|}{\partial\theta_i}\frac{\partial|\psi(\boldsymbol\theta)\rangle}{\partial\theta_j}]$ and $B_{i}=-\mathrm{Re}[\frac{\partial\langle\psi(\boldsymbol\theta)|}{\partial\theta_i}H|\psi(\boldsymbol\theta)\rangle]$. To realize the variational-based QITE with quantum circuits, we represent the parameterized state by a set of parameterized unitary gates $\{U_i(\theta_i)\}$, $|\psi(\boldsymbol\theta)\rangle=V(\boldsymbol\theta)|\psi_0\rangle=\prod_i U_i(\theta_i)|\psi_0\rangle$. By expanding the system Hamiltonian into the form of Pauli operators, $H = \sum_l h_l \sigma_l$, we can naturally write the derivative of each unitary gate in the form $\partial U_i(\theta_i)/ \partial\theta_i = \sum_k p_{k,i} U_i(\theta_i) \sigma_{k,i}$. The elements of the matrices $A, B$ can be rewritten as
\begin{eqnarray}
A_{ij} &=& \mathrm{Re}[\sum_{k,l}p_{k,i}^\ast p_{l,j}\langle\psi_0|W_{k,i}^\dag W_{l,j}|\psi_0\rangle]\no\\
B_{i} &=& -\mathrm{Re}[\sum_{k,l}p_{k,i}^\ast h_l \langle\psi_0|W_{k,i}^\dag \sigma_l V(\boldsymbol\theta)|\psi_0\rangle],
\label{eqs_004}
\end{eqnarray}
with $W_{k,i} = U_\gamma(\theta_\gamma) \cdots U_i(\theta_i)\sigma_{k,i} \cdots U_1(\theta_1)$. The real part of each expected value in the summation sign on the right side of Eq.~(\ref{eqs_004}) can be estimated by introducing an ancillary qubit. Without loss of generality, for any system operator $U_s$, the real part of its expected value under state $|\xi_s\rangle$ is calculated as
\begin{eqnarray}
\mathrm{Re}[e^{i\varphi}\langle\xi_s|U_s|\xi_s\rangle] = Tr(\rho_a Z_a),
\label{eqs_005}
\end{eqnarray}
which can be obtained by a Hadamard measurement of the ancillary qubit with its initial state prepared in $(|0_a\rangle+e^{i\varphi}|1_a\rangle)/\sqrt{2}$.

\section{Ansatzes for $\boldsymbol{\mathrm{H_2}}$ and $\boldsymbol{\mathrm{LiH}}$}\label{sec3}
The parameterized circuits $V(\boldsymbol\theta)$ mentioned in Sec.~\ref{sec2} can be designed with different configurations. Here we study $\mathrm{H_2}$ and LiH molecules using the UCC and hardware-efficient ansatzes. With two-electron orbitals in $\mathrm{H_2}$ and the spin freedom, the full mapping of $\mathrm{H_2}$ requires four qubits. Combining the quantum chemistry package OpenFermion~\cite{OpenFermion}, we perform a Bravyi-Kitaev (BK) transformation~\cite{SeeleyJCP12} on the second-quantization Hamiltonian of $\mathrm{H_2}$ under the STO-3G basis. We can find that the entanglement only occurs within two qubits. After averaging the remaining qubits, we obtain a Hamiltonian that contains at most two-qubit interactions, and the calculated results are close to the reference~\cite{MalleyPRX16}. With six-electron orbitals in LiH and the spin freedom, we need twelve qubits to simulate LiH. Because some orbitals contribute little to the bonding correlation, we further apply the active-space approximation to reduce the number of qubits required in the mapping process. Based on the natural orbital occupation number (NOON), we select three orbitals of LiH as the active space, while the NOON is obtained by the configuration interaction calculation with single and double excitations (CISD). After averaging the remaining qubits in LiH, we obtain a Hamiltonian that contains at most three-qubit interactions under the STO-6G basis~\cite{HempelPRX18}. The corresponding Hamiltonian data of LiH are presented in TABLE~\ref{LiH_data} (see Appendix~\ref{appC}).

UCC is a physical ansatz with low circuit depth, which usually starts from the Hartree-Fork molecular orbital wave function and works through the excited coupled-cluster operator $J$. UCC operator $V_\mathrm{UCC}=e^{J-J_\dag}$, in which $J$ can be converted into the form of Pauli matrices after the BK transformation. UCC is generally truncated to the single and double excitations ($J=J_1+J_2$), known as the UCCSD. With the Trotter-Suzuki decomposition, UCC operator $V_\mathrm{UCC}$ can be further transformed into a bunch of parameterized single-qubit and two-qubit gates.

For one $\mathcal{N}$-orbital system with $\mathcal{M}$ electrons, with Jordan-Wigner (JW) transformation~\cite{JW1928}, the Hartree-Fork state can be expressed as the direct product of single-qubit basis vectors $|\mathrm{HF}\rangle = |1\rangle^{\otimes \mathcal{M}}|0\rangle^{\otimes \mathcal{N}-\mathcal{M}}$. For H$_2$, the full Hartree-Fork state under JW form is $|\mathrm{HF}(\mathrm{H}_2)\rangle_{\mathrm{JW}} =|1100\rangle$, and the corresponding state under BK form is $|\mathrm{HF}(\mathrm{H}_2)\rangle_{\mathrm{BK}} =|1000\rangle$. The reduced Hartree-Fork state is further given by $|\mathrm{HF}(\mathrm{H}_2)\rangle =|10\rangle$. The corresponding UCCSD operator can be calculated through the second-order M$\o$ller-Plesset (MP2) method~\cite{Plesset1934}, the result of which only contains the double excitation operator $J_2$. Then the UCC operator is $V_\mathrm{UCC}(\mathrm{H_2},\theta)=e^{-i X_0 Y_1 \theta}$ in Pauli form, including only one variational parameter $\theta$. For LiH, the reduced Hartree-Fork state is given by $|\mathrm{HF} (\mathrm{LiH})\rangle=|100\rangle$. Its UCC operator is $V_\mathrm {UCC}(\mathrm{LiH},\boldsymbol\theta)=e^{-i X_0 Y_2 \theta_2}e^{-i X_0 Y_1 \theta_1}$, including two parameters $\theta_1$ and $\theta_2$. The ansatz state can be constructed from the UCC operator and Hartree-Fork state in the form of $V_\mathrm {UCC}(\boldsymbol\theta)| \mathrm {HF}\rangle$.

In addition to UCC, another feasible circuit is the compact compiling hardware-efficient ansatz, which can be more conveniently implemented in NISQ devices. Based on the qubit interaction existing in an actual quantum chip, the repeating unit of the hardware-efficient ansatz sequence is built from a layer of single-qubit rotated gates followed by a layer of two-qubit entangled gates. The number of repetitions in hardware-efficient ansatz is defined as the depth $D$. The depth has to be increased with the increased complexity of Hamiltonian, although a trade-off often exists between the depth and experimental fidelity. The hardware-efficient ansatz can be generally applied to different quantum systems but has more variational parameters than the UCC ansatz, which makes the practical experiment more resource-intensive. Both ansatzes have been applied to implement the variational-based QITE algorithm, with results shown in the following section.

\section{Results of two molecules}\label{sec4}

\subsection{Experimental Setup}\label{sec_040}

\begin{figure}[t]
\includegraphics{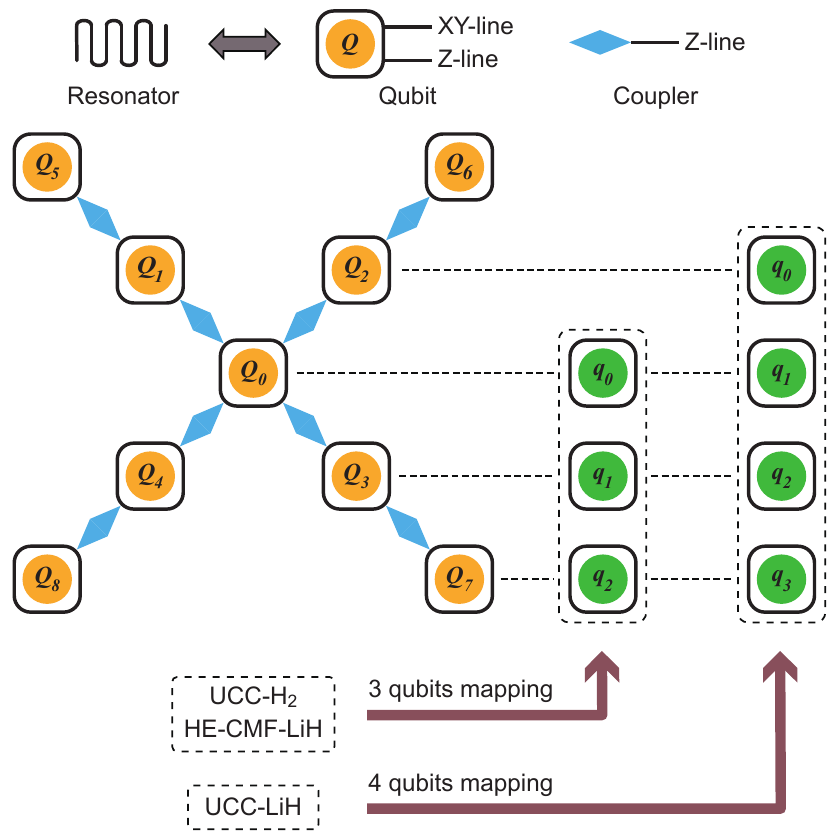}
\caption{Schematic diagram of the superconducting 9-qubit chip. Each qubit has the XY, and Z control lines to manipulate the two-level quantum state. The tunable coupler can be biased to change the total effective coupling strength between neighboring qubits. At the end of quantum manipulations, the qubit state can be simultaneously measured through the dispersive resonators. We mainly select three or four qubits from this chip and reordered the qubit numbers to implement the QITE circuit for $\mathrm{H_2}$ and LiH.}
\label{fig_n01}
\end{figure}

\begin{figure}[t]
\includegraphics{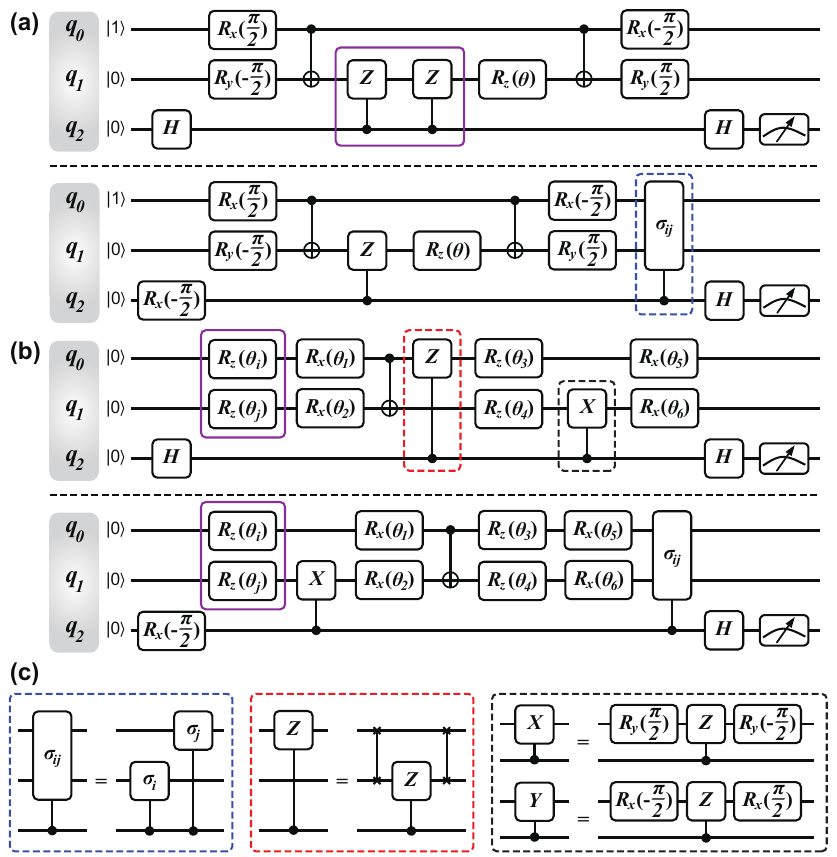}
\caption{QITE circuit with UCC and hardware-efficient ansatzes. (a) Quantum circuit for solving the hydrogen molecule based on UCC ansatz. Top: the circuit to estimate coefficient $A$, which is finally obtained by measuring the expected value of ancillary qubit $q_2$. For simplicity, the two consecutive CZ gates in the purple box can be omitted to de-entangle $q_2$. Bottom: the circuit to estimate different terms in B. The Pauli operator $\sigma_{ij}=\sigma_{i}\sigma_{j}$, representing a two-qubit interaction with $\sigma_{i,j}=I, X, Y, Z$. The initial guess is usually selected as a Hartree-Fork state $|\mathrm{HF} (\mathrm{H_2})\rangle=|10\rangle$. (b) Circuit for solving hydrogen or lithium hydride molecules based on the hardware-efficient ansatz. For example, the matrix element $A_{3,6}$ and $B_2$ can be estimated by the top and bottom circuits, respectively. The rotation gates $R_z(\theta_{i,j})$ in purple boxes can also be effectively omitted in the experiment. (c) The equivalent and practical realization of quantum gates in the dashed squares. }
\label{fig_n02}
\end{figure}

The variational-based QITE algorithm is implemented on a superconducting 9-qubit chip with tunable couplers~\cite{ChenPRL14, YanFeiPRApp18}. As shown in Fig.~\ref{fig_n01}, each transmon qubit is connected to its $XY/Z$ control lines and readout resonator~\cite{BarendsPRL13}. Through Z-control lines, the qubits are initially $DC$-biased at their idle point, and the couplers are tuned as far as possible to suppress the $ZZ$ interaction between neighboring qubits~\cite{MundadaPRApp,LuyanSun20,ZhongchuPRL20}. The microwave pulse sequence consisting of single- and two-qubit gates is generated by the arbitrary waveform generator (AWG) boards~\cite{ZhangMengyu}, and the resulting qubit state information is encoded into the resonator in the dispersion region. The readout signal can be further processed by cryogenic low-noise amplifiers, and finally transmitted to the data acquisition (DAQ) board for demodulation~\cite{ZhangMengyu}.

In this experiment, we mainly focus on three or four qubits $Q_0, Q_3, Q_7$ ($Q_2, Q_0, Q_3, Q_7$) which are closely adjacent in space, and relabel the qubit number as $q_0,q_1,q_2$ ($q_0,q_1,q_2,q_3$) to implement the QITE algorithm. For $q_0$ to $q_3$, the average energy relaxation time at their idle points is $\overline{T}_{ 1} \approx 28.7~\mu$s, and the ramsey dephasing time is $\overline{T}_{2} \approx 26.5~\mu$s. The average readout fidelity of the ground state $|g\rangle$ and the excited state $|e\rangle$ are $F_g \approx 97.9\%$ and $F_e \approx 93.2\%$, and the readout correction of each qubit is embedded in the experimental program. By a fine calibration of the pulse distortion, the crosstalk, and the timing deviation in our hardware, the gate error can be optimized to a relatively low level, with the average single-qubit gate error $e_{sq} \approx 0.1 \%$ and two-qubit CZ gate error $e_{cz} \approx 1.1 \%$. Next, we will present the ground state simulation of $\mathrm{H_2}$ and LiH with two ansatzes on this 3- and 4-qubit system.

\subsection{Hydrogen Molecule}\label{sec_041}

\begin{figure}[t]
\includegraphics{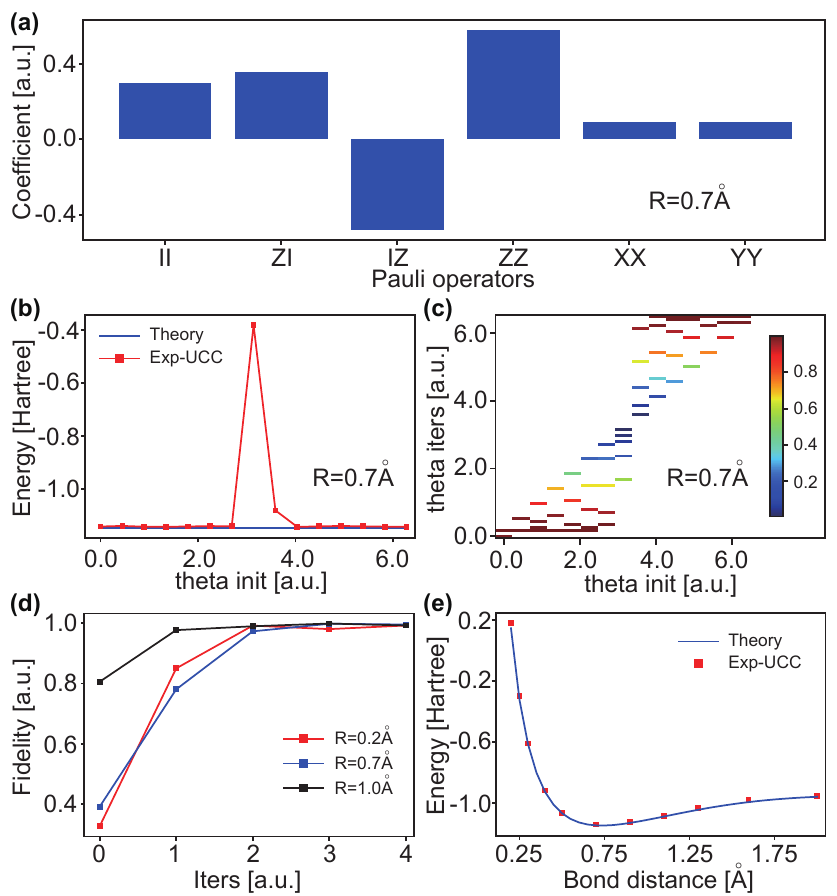}
\caption{QITE result with UCC ansatz for solving the ground state of $\mathrm{H_2}$. (a) Theoretical Pauli decomposition of $\mathrm{H_2}$ near the equilibrium point $R=0.7\ \mathring{A}$. (b) The final energy (in Hartree) after the same number of iterations (with $l=4$) versus different initial attempts of $\theta^{(0)}$. (c) The detailed iteration results. For each different initial  $\theta^{(0)}$, four sequentially $\theta^{(l)}$ are shown by small rectangles at updated angles, with the color representing measured state fidelity. Most initial $\theta^{(0)}$ converges to a similar value with high fidelity. (d) With the initial guess $\theta^{(0)}=2.0$, three optimization curves of state fidelity are shown for demonstration, including the $R=0.2\ \mathring{A}$ (red) at one Coulomb repulsion point, the $R=0.7\ \mathring{A}$ (blue) near the equilibrium point and the $R=2.0\ \mathring{A}$ (black) at one Coulomb attractive point. (e) The experimental ground state energy at different bond distances is consistent with the theoretical blue curve.}
\label{fig_n03}
\end{figure}

Any unitary operator can be theoretically split into a series of single- and two-qubit gates~\cite{LloydPRL95}. To realize the UCC operator $V_\mathrm{UCC}(\mathrm{H_2},\theta)=e^{-i X_0 Y_1 \theta}$ of H$_2$ in a quantum computing system, we can rewrite the equivalent gate circuit as
\begin{eqnarray}
V_\mathrm{UCC}(\mathrm{H_2},\theta)&=&R_{x}^{q_0}(-\frac{\pi}{2})R_{y}^{q_1}(\frac{\pi}{2}) \mathrm{CNOT}(q_0, q_1) R_{z}^{q_1}(2\theta)\no\\
&&\mathrm{CNOT}(q_0, q_1) R_{x}^{q_0}(\frac{\pi}{2}) R_{y}^{q_1}(-\frac{\pi}{2}),
\label{eqs_006}
\end{eqnarray}
in which $R_{n=x,y,z}^{q_i}(\alpha)=e^{-i\frac{\alpha}{2}\sigma_n}$ is the rotation gate of single qubit $q_i$ and $\mathrm{CNOT}(q_0, q_1)$ is the entangled CNOT gate with control qubit $q_0$ and target qubit $q_1$. For convenience, we reset the parameter $2\theta$ in $R_{z}$ to $\theta$. According to Eq.~(\ref{eqs_006}), we can convert the derivative of the UCC operator to the derivative of rotation $Z$-gate, $\frac{\partial}{\partial \theta} V_\mathrm{UCC}(\mathrm{H_2},\theta) \Rightarrow \frac{\partial}{\partial \theta} R_{z}(\theta)=-\frac{i}{2}Z R_{z}(\theta)$, and there is only one variational parameter $\theta$ that needs to be determined. The detailed QITE circuit with UCC ansatz for $A$ and $B$ discussed in Sec.~\ref{sec2} is displayed in Fig.~\ref{fig_n02}(a). When introducing the ancillary qubit $q_2$, each item in $A$ and $B$ can be estimated by Eq.~(\ref{eqs_005}). For the control-$\sigma$ term in Fig.~\ref{fig_n02}, we can experimentally realize it by the gate decomposition $c\mbox{-}\sigma_{i,j}=c\mbox{-}\sigma_j \cdot c\mbox{-}\sigma_i$.

We first study the ground state of $\mathrm{H}_2$ when the bond distance is close to the equilibrium point. Figure~\ref{fig_n03}(a) shows the theoretical Pauli expansion form of the Hamiltonian $H(\mathrm{H_2})$ with $R=0.7\ \mathring{A}$. Starting with different variational parameters $\theta^{(0)}$, the initial guess state evolves as $|\psi(\theta^{(0)})\rangle = V_\mathrm{UCC}(\mathrm{H_2},\theta^{(0)})|10\rangle$, with the density matrix $\rho(\theta^{(0)})$ experimentally reconstructed from the quantum-state-tomography (QST)~\cite{DanielPRA01}. By empirically setting the imaginary time step to $d\tau = 0.8/(h_Z l)$, with the average $z$-component coefficient $h_Z=(h_{ZI}+h_{IZ})/2$ and the total number of iterations $l=4$, we apply the QITE circuit to measure the coefficients $A^{(0)}$ and $B^{(0)}$. With Euler's method $\theta^{(1)}=\theta^{(0)}+{A^{(0)}}^{-1}B^{(0)} d\tau$, the first updated state $\rho(\theta^{(1)})$ can be measured after one iteration. Repeating the same two circuits $l$ times, we can obtain the final output state $\rho(\theta^{(4)})$, and the state energy $E_0(\theta^{(4)})=\mathrm{Tr}(\rho(\theta^{(4)})H(\mathrm{H}_2))$ can be further determined. Figure~\ref{fig_n03}(b) and (c) give the visual data distribution with $\theta^{(0)}$ varying from $0$ to $2 \pi$. Except for the state at $\theta^{(0)} \approx \pi$ that has a large orthogonality with the ground state, most initial guesses of $\theta^{(0)}$ can converge to one with the lowest energy. With $\theta^{(0)}=2.0$, we optimize the state fidelity $F_{\rho(\theta^{(i)})}=\mathrm{Tr}(\rho_{g,\mathrm{th}}\rho(\theta^{(i)}))$ from $F_{\rho(\theta^{(0)})} \approx 39.2\%$ to $F_{\rho(\theta^{(4)})} \approx 99.5\%$. For other values of $R$ in the Coulomb attraction and repulsion regions, we apply the same $\theta^{(0)}$ and update method. Figure~\ref{fig_n03}(d) shows the experimental result for three different bond distances. The state fidelity quickly converges to a value close to 1.0 after 4 iterations, which proves the effectiveness and efficiency of this variational-based QITE. In Fig.~\ref{fig_n03}(e), experimental results of the ground state energy as a function of $R$ also agree well with the theoretical curve. The core steps for solving the ground state of $\mathrm{H_2}$ with UCC ansatz are listed in the Algorithm 1.

\begin{figure}[t]
\includegraphics{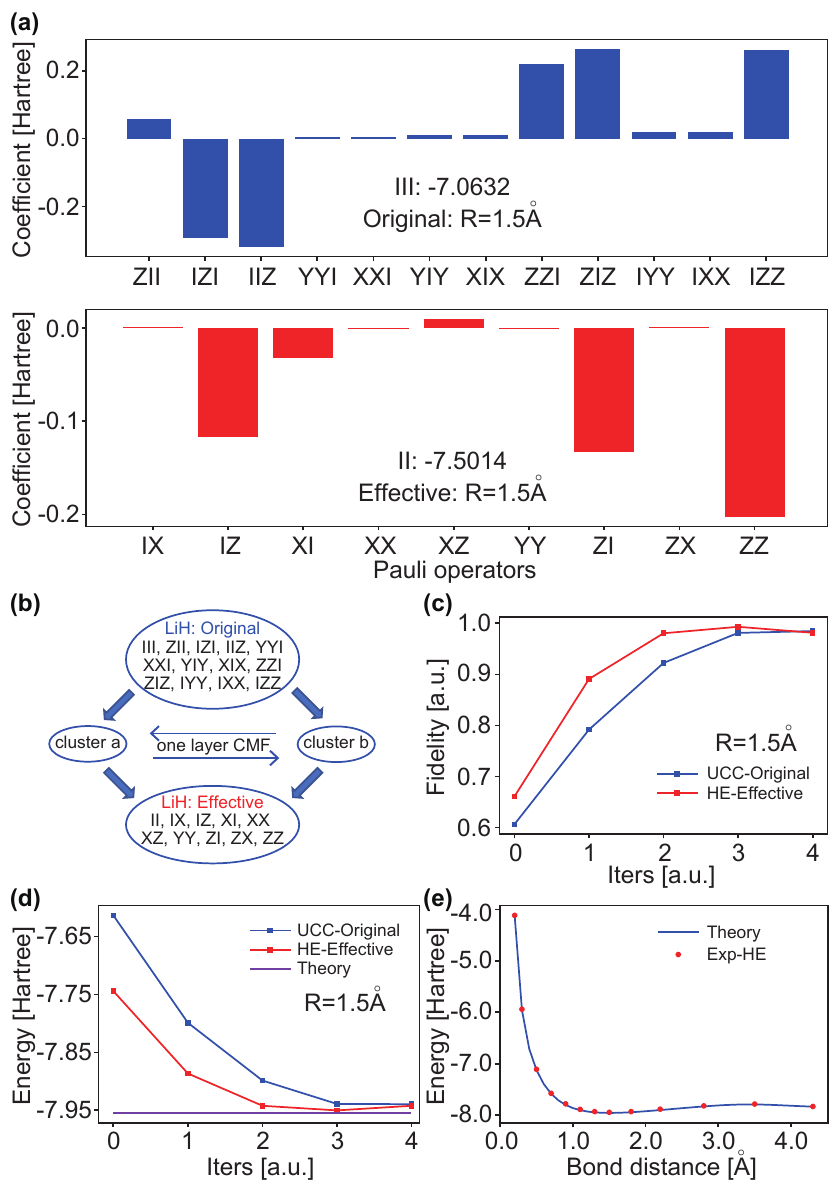}
\caption{QITE result with hardware-efficient ansatz for solving the ground state of LiH. (a) Theoretical Pauli decomposition of the original and CMF-reduced Hamiltonian $\mathrm{LiH}$ at the equilibrium point $R=1.5\ \mathring{A}$. The $\mathrm{III} (\mathrm{II})$ term is relatively large and only given by a numerical value. For other bond distances, the Hamiltonian data are listed in TABLE~\ref{LiH_data} in detail. (b) A schematic diagram of the CMF method for LiH, with the effective Hamiltonian reconstructed on a new set of bases states. (c) The ground state fidelity at $R=1.5\ \mathring{A}$ is optimized based on both the UCC and hardware-efficient ansatzes. (d) The convergence curves of ground state energy corresponding to (c), together with the theoretical result of energy. (e) The experimental CMF data with hardware-efficient ansatz at different bond distances, which is well consistent with the theoretical curve. }
\label{fig_n04}
\end{figure}

\begin{algorithm}[H]
\caption{QITE with UCC ansatz for solving $\mathrm{H_2}$}
\begin{algorithmic}[1]
\State Select $l$
\Comment{number of iterations}
\State $h_z = (h_{ZI}+h_{IZ})/2$
\Comment{average z component}
\State $d\tau \propto 1/(h_z l)$
\Comment{imaginary time interval}
\State $T = l d\tau$
\Comment{total time}
\For {$r$ in $\{r\}$}
    \State $H=H(r)$
    \Comment{Hamiltonian at bond distance $r$}
    \State Start with $\theta(t) = \theta^{(0)}$
    \For {$t$ in $[0,dt,...,T]$}
        \State Estimate $A(t)$ with the top circuit in Fig.~\ref{fig_n02}(a)
        \State Estimate $B(t)$ with the bottom circuit in Fig.~\ref{fig_n02}(a)
        \State Update $\theta(t) = \theta(t) + (B(t)/A(t))d\tau$
        \State Check the state $\rho(\theta)$
        \Comment{QST}
        \State Calculate energy $E_g=Tr(\rho(\theta) H)$
    \EndFor
    \State {Record the final state $\rho_r$ and energy $E_{g,r}$}
\EndFor
\State {Get the potential energy curve}
\end{algorithmic}
In lines 12 and 13 of the algorithm, we do not necessarily need to measure each intermediate state; we only care about the final result in the last step.
\end{algorithm}
\label{alg_1}

The hardware-efficient ansatz can also solve the ground state energy of $\mathrm{H_2}$. However, in order to simplify the experiment, we directly apply this ansatz to the more complicated LiH molecule.

\subsection{Lithium Hydride Molecule}\label{sec_042}

For efficient use of qubit resources, we first implement QITE with the hardware-efficient ansatz to solve the ground state of LiH in a 3-qubit system. The original LiH Hamiltonian contains three-qubit interactions, then there will be four qubits involved in the QITE circuit. In the top part of Fig.~\ref{fig_n04}(a), we plot the theoretical Hamiltonian near the equilibrium point $R=1.5\ \mathring{A}$ in Pauli form. To reduce the qubit consumption or circuit complexity, we introduce the one-layer CMF method to obtain an effective Hamiltonian of LiH in a smaller system (Fig.~\ref{fig_n04}(b)).

In the CMF method, the Hamiltonian $H(\mathrm{LiH})$ can be divided into two subsystems $a(2)$ and $b(1)$ including two qubits and one qubit, respectively. We suppose that the initial state of system $b$ is $\rho_b^{(0)}=\frac{1}{2}(I_b+X_b)$, in which $I$ is the identity matrix. Then the reduced Hamiltonian of the system $a$ under $\rho_b^{(0)}$ is $H_a^{(0)}(\mathrm{LiH})=\mathrm{Tr}_b(I_a\rho_b^{(0)}H)$, and the two lowest eigenstates of $H_a^{(0)}(\mathrm{LiH})$ are $|\psi_{a(g)}^{(0)}\rangle$ and $|\psi_{a(e)}^{(0)}\rangle$. Under the two eigenstates of $a$, the reduced Hamiltonian of $b$ can be expressed as $H_b^{(0)}(\mathrm{LiH})=\mathrm{Tr}_a(\rho_{a(g,e)}^{(0)}I_b H)$, and the corresponding four eigenstates are $|\psi_{b_g(a_g)}^{(1)}\rangle$, $|\psi_{b_e(a_g)}^{(1)}\rangle$, $|\psi_{b_g(a_e)}^{(1)}\rangle$, and$|\psi_{b_e(a_e)}^{(1)}\rangle$. Through further iterations, the reduced Hamiltonian $H_a^{(1)}(\mathrm{LiH})=\mathrm{Tr}_b(I_a\rho_b^{(1)}H)$ have eight relevant eigenstates $\{|\psi_a^{(1)}(j)\rangle,j=0,1,\cdots,7\}$. We terminate at this step, and artificially select four states in $\{|\psi_a^{(1)}\rangle\}$ to make a direct product with $\{|\psi_b^{(1)}\rangle\}$, finally get a set of quantum states $\{|\psi_{a,b}^{(1)}(j)\rangle,j=0,1,2,3\}$. With Schmidt orthogonalization, a new set of bases $\{|\psi_\mathrm{eff}(j)\rangle,j=0,1,2,3\}$ can be obtained. The effective Hamiltonian $H_{\mathrm{eff}}(\mathrm{LiH})=\sum_{i,j}\langle\psi_\mathrm{eff}(i)|H(\mathrm{LiH})|\psi_\mathrm{eff}(j)\rangle|i\rangle \langle j|$ provide a good estimation of the original system. The lower part of Fig.~\ref{fig_n04}(a) shows the theoretical result of $H_\mathrm{eff}(\mathrm{LiH})$ corresponding to the value of $R=1.5\ \mathring{A}$.

The choice of such a Hamiltonian can achieve a ground state fidelity of $F_{|\psi_g\rangle}=|\langle\psi_{g,\mathrm{th}}|\psi_{g,\mathrm{sim}}\rangle|^2 > 99.9\%$ in our simulation. Similarly, we can also divide the system into the form of $a(1)$ and $b(2)$, or use the combination of different groups to select a more effective Hamiltonian. In practice, these choices will change according to the system characteristics or experimental conditions. Next, we will employ the hardware-efficient ansatz in Fig.~\ref{fig_n02}(b) to solve the effective Hamiltonian.

The simplified hardware-efficient operator of LiH can be written as
\begin{eqnarray}
V_\mathrm{HE}(\mathrm{LiH},\boldsymbol\theta)&=&R_{x}^{q_1}(\theta_6)R_{x}^{q_0}(\theta_5)R_{z}^{q_1}(\theta_4) R_{z}^{q_0}(\theta_3)\no\\
&&\mathrm{CNOT}(q_0, q_1) R_{x}^{q_1}(\theta_2) R_{x}^{q_0}(\theta_1),
\label{eqs_007}
\end{eqnarray}
with a depth setting of $D=1$. The corresponding circuit contains six variational parameters. Each element in $A_{6\times6}$ and $B_{6\times1}$ can be obtained by measuring the state of the ancillary qubit. Due to the fact that $A_{ij} = A_{ji}$, all transposed elements ($A_{ji}$) have been omitted in our experiment. With an initial guess list $\boldsymbol{\theta}^{(0)}=[\theta_1^{(0)},\theta_2^{(0)},\theta_3^{(0)},\theta_4^{(0)},\theta_5^{(0)},\theta_6^{(0)}]=[0.5]*6$ and the updated procedure in Algorithm 2, we can optimize the quantum state with fidelity from $F_{\rho(\boldsymbol{\theta}^{(0)})} \approx 66.1\%$ to $F_{\rho(\boldsymbol{\theta}^{(4)})} \approx 98.2\%$ at $R=1.5\ \mathring{A}$.

\begin{algorithm}[H]
\caption{QITE with HE ansatz for solving $\mathrm{LiH}$}
\begin{algorithmic}[1]
\State Select $l$
\Comment{number of iterations}
\State $h_z = (h_{ZII}+h_{IZI}+h_{IIZ})/3$
\Comment{average z component}
\State $d\tau \propto 1/(h_z l)$
\Comment{imaginary time interval}
\State $T = l d\tau$
\Comment{total time}
\For {$r$ in $\{r\}$}
    \State $H=H(r)$
    \Comment{Hamiltonian at bond distance $r$}
    \State $H \rightarrow H_{eff}$
    \Comment{effective Hamiltonian with CMF}
    \State Start with $\boldsymbol{\theta(t)} = \boldsymbol{\theta^{(0)}}$
    \For {$t$ in $[0,dt,...,T]$}
        \State Estimate $A(t)$ with the top circuit in Fig.~\ref{fig_n02}(b)
        \State Estimate $B(t)$ with the bottom circuit in Fig.~\ref{fig_n02}(b)
        \State Update $\boldsymbol{\theta(t)} = \boldsymbol{\theta(t)} + A^{-1}(t) B(t) d\tau$
        \State Check the state $\rho(\boldsymbol{\theta})$
        \Comment{QST}
        \State Transform base vector $\rho(\boldsymbol{\theta}) \rightarrow \rho^{'}(\boldsymbol{\theta})$
        \State Calculate energy $E_g=Tr(\rho^{'}(\boldsymbol{\theta}) H)$
    \EndFor
    \State {Record the final state $\rho_r$ and energy $E_{g,r}$}
\EndFor
\State {Get the potential energy curve}
\end{algorithmic}
\end{algorithm}
\label{alg_2}

For comparison, we also implement the QITE circuit based on UCC ansatz in a 4-qubit system. The detailed QITE circuit is given in Appendix~\ref{appB} and Fig.~\ref{fig_n05}. With $\boldsymbol{\tilde{\theta}^{(0)}}=[1.0]*2$, the state fidelity at the the equilibrium point $R=1.5\ \mathring{A}$ is updated and optimized from $F_{\rho(\boldsymbol{\tilde{\theta}}^{(0)})} \approx 60.1\%$ to $F_{\rho(\boldsymbol{\tilde{\theta}}^{(4)})} \approx 98.5\%$. In Fig.~\ref{fig_n04}(c) and (d), both the state fidelity and the ground state energy quickly converges to optimal values with high fidelity. Both results are comparable to the convergence curves using hardware-efficient ansatz, while the slight difference may be due to the difference in system dimensions and circuit details.

\section{Conclusion}\label{sec6}
In this paper, we mainly study the variational-based QITE algorithm under UCC and hardware-efficient ansatzes.  We experimentally solve the ground states of H$_2$ and LiH molecules by measuring the ancillary qubit on the superconducting qubit platform. Compared to the traditional VQE, optimization of the QITE algorithm is guided by the physical principle that the quantum state can rapidly converge to the ground state with all excited components filtered by the exponential decay. The experimental results show that the optimization can be easily realized with a few iterations. However, for a relatively complicated system such as the LiH molecule, the required qubit number is comparatively larger. For the hardware-efficient with more gate parameters, the actual experiment is quite resource-intensive. We practically apply the CMF method to obtain an effective Hamiltonian in a compressed subspace. For the excited state energy, we propose a method in Appendix~\ref{appA} to simulate it with existing algorithms such as the variational-based QITE circuit. The system Hamiltonian can be rearranged through the measured ground state and the estimated maximum eigenenergy, from which the excited states can be transformed into the ground state problem in the rearranged Hamiltonian.

In this preliminary experiment to verify the QITE algorithm, we simulate two molecules that do not have a high enough dimension. It is still very challenging to verify the feasibility of this hybrid scheme for more general models. Although the CMF method can reduce the number of qubits required for simulation, the effective Hamiltonian of a more complex system normally requires a certain time cost for the artificial reconstruction. Then the hybrid QITE algorithm includes the time-for-space consumption in the process. Nevertheless, it provides a possibility for simulating a larger Hamiltonian in NISQ systems with a smaller qubit system. In the future, we will further explore and improve the ground state simulation with QITE and CMF algorithms for different physical and chemical models.

\section*{Acknowledgements}
The work reported here was supported by the National Natural Science Foundation of China (Grants No. 12074336, No. 11934010), the National Key Research and Development Program of China (Grant No. 2019YFA0308602), the Fundamental Research Funds for the Central Universities in China (2020XZZX002-01). Y.Y. acknowledge the funding support from Tencent Corporation.

\appendix

\section{Excited States}\label{appA}
For further solving the excited state of the system, we propose a general method to transform the excited state into the ground state with the Gershgorin circle theorem~\cite{Gerschgorin}. Originally proposed by S. A. Gershgorin in 1931, the Gershgorin circle theorem in mathematics is initially used to limit the range of eigenvalues. Here we apply it to the energy spectrum of a target Hamiltonian. According to the Gershgorin circle theorem, the eigenvalues of a square matrix exist in the union set of a series of Gershgorin discs. For the Hamiltonian $H_{N \times N}$, each eigenenergy lies within at least one of the closed circles $E_n \in \{D(H_{i,i}, R_i)\} (0 \leq i,n \leq N-1$), with the center $H_{i,i}$ and radius $R_i=\sum_{j \neq i}^{N-1} |H_{i,j(j,i)}|$. Then we can estimate the bound of the largest eigenvalue $E_{N-1}<E_{max}$ and rearrange the energy levels as $E_0,E_1,\cdots,E_{N-1} \to E_1,\cdots,E_{N-1},E_0+(E_{max}-E_0)$. The system Hamiltonian is transformed from $H=\sum_{n=0}^{N-1}E_n|\phi_n\rangle\langle\phi_n|$ to $H'=E_{max}|\phi_0\rangle\langle\phi_0|+\sum_{n=1}^{N-1}E_n|\phi_n\rangle\langle\phi_n|$. The first excited state of the original system $H$ becomes the ground state of $H'$, which can be solved by the ground-state algorithm such as QITE. The same process can be performed for further excited states with higher energy levels. However, because the rearranged Hamiltonian depends on the previously measured eigenstate, the error may be increasingly accumulated for higher energy levels.

\section{UCC Ansatz of LiH}\label{appB}
\begin{figure*}[htb]
\includegraphics{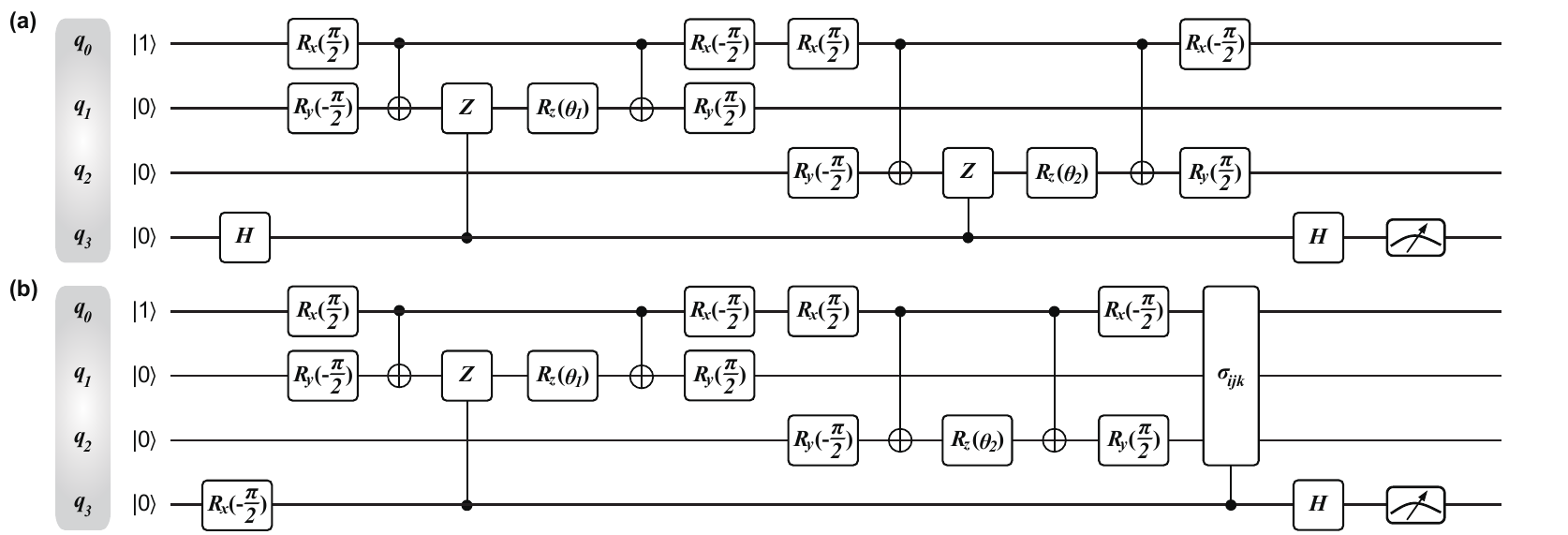}
\caption{QITE circuit with UCC ansatz to solve LiH. (a) Estimation of the matrix element $A_{1,2}$. In theory, the off-diagonal elements of $A$ are zero. (b) Estimation of the $B_1$ with the three body Pauli operator $\sigma_{ijk}=\sigma_{i}\sigma_{j}\sigma_{k}$. In both circuits, the qubits are initially prepared in the Hartree-Fock state $|\mathrm{HF} (\mathrm{LiH})\rangle=|100\rangle$.}
\label{fig_n05}
\end{figure*}

The UCC operator of LiH mentioned before is $V_\mathrm {UCC}(\mathrm{LiH},\boldsymbol{\theta})=e^{-i X_0 Y_2 \theta_2}e^{-i X_0 Y_1 \theta_1}$, with $\boldsymbol{{\theta}}=[\theta_1,\theta_2]$. We can write the equivalent gate circuit as
\begin{eqnarray}
V_\mathrm{UCC}(\mathrm{LiH},\boldsymbol{\theta})&=&R_{x}^{q_0}(-\frac{\pi}{2}) R_{y}^{q_2}(\frac{\pi}{2}) \mathrm{CNOT}(q_0, q_2) R_{z}^{q_2}(2\theta_2)\no\\
&&\mathrm{CNOT}(q_0, q_2) R_{x}^{q_0}(\frac{\pi}{2}) R_{y}^{q_2}(-\frac{\pi}{2})\no\\
&&R_{x}^{q_0}(-\frac{\pi}{2}) R_{y}^{q_1}(\frac{\pi}{2}) \mathrm{CNOT}(q_0, q_1) R_{z}^{q_1}(2\theta_1)\no\\
&&\mathrm{CNOT}(q_0, q_1) R_{x}^{q_0}(\frac{\pi}{2}) R_{y}^{q_1}(-\frac{\pi}{2}).
\label{eqs_B01}
\end{eqnarray}
Combined with the ancillary qubit, the QITE circuits for solving $A$ and $B$ can be mapped to a 4-qubit system. The example circuits for solving the matrix elements of $A$ and $B$ are shown in Fig.~\ref{fig_n05}.

\section{LiH Data}\label{appC}
With BK transformation on the second quantization Hamiltonian under the STO-6G basis, we give the results of the LiH Hamiltonian calculated by Google's open-source quantum chemistry package OpenFermion. The bond distance for all values of $R$ ranges from $0.1\ \mathring{A}$ to $5.0\ \mathring{A}$. The Hamiltonian at each value of $R$ can be decomposed into 13 Pauli interaction terms. The specific data is presented in table~\ref{LiH_data}.

\begin{table*}[htb]
\caption{The Hamiltonian coefficients for LiH at each value of $R$.}
\begin{threeparttable}
\begin{tabular}{cp{1.18cm}<{\centering}p{1.18cm}<{\centering}p{1.18cm}<{\centering}p{1.18cm}<{\centering}p{1.18cm}<{\centering}p{1.18cm}<{\centering}p{1.18cm}<{\centering}p{1.18cm}<{\centering}p{1.18cm}<{\centering}p{1.18cm}<{\centering}<{\centering}p{1.18cm}<{\centering}p{1.18cm}<{\centering}p{1.18cm}<{\centering}p{1.18cm}}
\hline
\hline
$R$& III & ZII & IZI & IIZ & YYI & XXI & YIY & XIX & ZZI & ZIZ & IYY & IXX & IZZ \\ \hline
0.1 & 3.3063 & -0.0694 & -0.2029 & -0.2048 & 0.0314 & 0.0314 & 0.0319 & 0.0319 & 0.2648 & 0.2655 & 0.0086 & 0.0086 & 0.2713 \\
0.2 & -3.5316 & -0.0740 & -0.1932 & -0.2023 & 0.0285 & 0.0285 & 0.0307 & 0.0307 & 0.2593 & 0.2614 & 0.0092 & 0.0092 & 0.2723 \\
0.3 & -5.3654 & -0.0775 & -0.1883 & -0.2086 & 0.0238 & 0.0238 & 0.0290 & 0.0290 & 0.2580 & 0.2640 & 0.0104 & 0.0104 & 0.2729 \\
0.4 & -6.0969 & -0.0720 & -0.1917 & -0.2212 & 0.0150 & 0.0150 & 0.0253 & 0.0253 & 0.2558 & 0.2737 & 0.0137 & 0.0137 & 0.2700 \\
0.5 & -6.4555 & -0.0527 & -0.2028 & -0.2357 & 0.0065 & 0.0065 & 0.0195 & 0.0195 & 0.2533 & 0.2867 & 0.0188 & 0.0188 & 0.2636 \\
0.6 & -6.6627 & -0.0270 & -0.2172 & -0.2498 & 0.0049 & 0.0049 & 0.0164 & 0.0164 & 0.2551 & 0.2940 & 0.0213 & 0.0213 & 0.2610 \\
0.7 & -6.7927 & -0.0001 & -0.2332 & -0.2622 & 0.0046 & 0.0046 & 0.0151 & 0.0151 & 0.2547 & 0.2960 & 0.0218 & 0.0218 & 0.2608 \\
0.8 & -6.8791 & 0.0231 & -0.2476 & -0.2725 & 0.0045 & 0.0045 & 0.0143 & 0.0143 & 0.2516 & 0.2948 & 0.0217 & 0.0217 & 0.2613 \\
0.9 & -6.9390 & 0.0406 & -0.2594 & -0.2811 & 0.0044 & 0.0044 & 0.0138 & 0.0138 & 0.2470 & 0.2916 & 0.0215 & 0.0215 & 0.2617 \\
1.0 & -6.9814 & 0.0525 & -0.2686 & -0.2886 & 0.0045 & 0.0045 & 0.0135 & 0.0135 & 0.2420 & 0.2874 & 0.0212 & 0.0212 & 0.2620 \\
1.1 & -7.0115 & 0.0595 & -0.2758 & -0.2954 & 0.0046 & 0.0046 & 0.0131 & 0.0131 & 0.2370 & 0.2827 & 0.0210 & 0.0210 & 0.2622 \\
1.2 & -7.0327 & 0.0628 & -0.2813 & -0.3016 & 0.0048 & 0.0048 & 0.0128 & 0.0128 & 0.2323 & 0.2779 & 0.0208 & 0.0208 & 0.2623 \\
1.3 & -7.0473 & 0.0634 & -0.2857 & -0.3075 & 0.0050 & 0.0050 & 0.0124 & 0.0124 & 0.2280 & 0.2730 & 0.0207 & 0.0207 & 0.2623 \\
1.4 & -7.0571 & 0.0620 & -0.2890 & -0.3129 & 0.0053 & 0.0053 & 0.0121 & 0.0121 & 0.2239 & 0.2681 & 0.0207 & 0.0207 & 0.2622 \\
1.5 & -7.0632 & 0.0593 & -0.2916 & -0.3179 & 0.0057 & 0.0057 & 0.0118 & 0.0118 & 0.2201 & 0.2633 & 0.0207 & 0.0207 & 0.2620 \\
1.6 & -7.0667 & 0.0558 & -0.2935 & -0.3225 & 0.0061 & 0.0061 & 0.0115 & 0.0115 & 0.2166 & 0.2585 & 0.0207 & 0.0207 & 0.2618 \\
1.7 & -7.0683 & 0.0516 & -0.2947 & -0.3267 & 0.0066 & 0.0066 & 0.0112 & 0.0112 & 0.2133 & 0.2539 & 0.0207 & 0.0207 & 0.2616 \\
1.8 & -7.0685 & 0.0470 & -0.2954 & -0.3305 & 0.0072 & 0.0072 & 0.0110 & 0.0110 & 0.2102 & 0.2492 & 0.0208 & 0.0208 & 0.2612 \\
1.9 & -7.0678 & 0.0421 & -0.2956 & -0.3339 & 0.0079 & 0.0079 & 0.0108 & 0.0108 & 0.2074 & 0.2447 & 0.0208 & 0.0208 & 0.2608 \\
2.0 & -7.0664 & 0.0370 & -0.2953 & -0.3370 & 0.0087 & 0.0087 & 0.0106 & 0.0106 & 0.2047 & 0.2403 & 0.0208 & 0.0208 & 0.2603 \\
2.1 & -7.0646 & 0.0317 & -0.2945 & -0.3397 & 0.0096 & 0.0096 & 0.0104 & 0.0104 & 0.2024 & 0.2360 & 0.0208 & 0.0208 & 0.2597 \\
2.2 & -7.0628 & 0.0263 & -0.2932 & -0.3420 & 0.0107 & 0.0107 & 0.0103 & 0.0103 & 0.2002 & 0.2318 & 0.0208 & 0.0208 & 0.2591 \\
2.3 & -7.0610 & 0.0208 & -0.2913 & -0.3440 & 0.0119 & 0.0119 & 0.0102 & 0.0102 & 0.1984 & 0.2278 & 0.0208 & 0.0208 & 0.2583 \\
2.4 & -7.0594 & 0.0152 & -0.2888 & -0.3457 & 0.0134 & 0.0134 & 0.0102 & 0.0102 & 0.1969 & 0.2239 & 0.0208 & 0.0208 & 0.2574 \\
2.5 & -7.0582 & 0.0094 & -0.2857 & -0.3470 & 0.0152 & 0.0152 & 0.0102 & 0.0102 & 0.1957 & 0.2202 & 0.0208 & 0.0208 & 0.2563 \\
2.6 & -7.0576 & 0.0034 & -0.2819 & -0.3481 & 0.0172 & 0.0172 & 0.0102 & 0.0102 & 0.1948 & 0.2167 & 0.0208 & 0.0208 & 0.2550 \\
2.7 & -7.0576 & -0.0027 & -0.2773 & -0.3487 & 0.0196 & 0.0196 & 0.0103 & 0.0103 & 0.1944 & 0.2133 & 0.0207 & 0.0207 & 0.2535 \\
2.8 & -7.0584 & -0.0090 & -0.2719 & -0.3491 & 0.0223 & 0.0223 & 0.0104 & 0.0104 & 0.1943 & 0.2101 & 0.0206 & 0.0206 & 0.2518 \\
2.9 & -7.0601 & -0.0155 & -0.2656 & -0.3490 & 0.0255 & 0.0255 & 0.0105 & 0.0105 & 0.1946 & 0.2071 & 0.0205 & 0.0205 & 0.2498 \\
3.0 & -7.0626 & -0.0221 & -0.2585 & -0.3486 & 0.0291 & 0.0291 & 0.0107 & 0.0107 & 0.1953 & 0.2043 & 0.0204 & 0.0204 & 0.2474 \\
3.1 & -7.0660 & -0.0288 & -0.2507 & -0.3478 & 0.0330 & 0.0330 & 0.0109 & 0.0109 & 0.1964 & 0.2016 & 0.0203 & 0.0203 & 0.2448 \\
3.2 & -7.0702 & -0.0356 & -0.2422 & -0.3466 & 0.0374 & 0.0374 & 0.0110 & 0.0110 & 0.1978 & 0.1991 & 0.0201 & 0.0201 & 0.2418 \\
3.3 & -7.0750 & -0.0423 & -0.2332 & -0.3451 & 0.0421 & 0.0421 & 0.0112 & 0.0112 & 0.1995 & 0.1967 & 0.0200 & 0.0200 & 0.2386 \\
3.4 & -7.0801 & -0.0490 & -0.2241 & -0.3433 & 0.0471 & 0.0471 & 0.0114 & 0.0114 & 0.2013 & 0.1945 & 0.0198 & 0.0198 & 0.2351 \\
3.5 & -7.0855 & -0.0554 & -0.2149 & -0.3412 & 0.0522 & 0.0522 & 0.0116 & 0.0116 & 0.2033 & 0.1924 & 0.0196 & 0.0196 & 0.2315 \\
3.6 & -7.0908 & -0.0616 & -0.2060 & -0.3390 & 0.0574 & 0.0574 & 0.0118 & 0.0118 & 0.2051 & 0.1904 & 0.0194 & 0.0194 & 0.2277 \\
3.7 & -7.0959 & -0.0673 & -0.1976 & -0.3368 & 0.0624 & 0.0624 & 0.0120 & 0.0120 & 0.2069 & 0.1885 & 0.0192 & 0.0192 & 0.2239 \\
3.8 & -7.1005 & -0.0726 & -0.1897 & -0.3346 & 0.0673 & 0.0673 & 0.0121 & 0.0121 & 0.2084 & 0.1867 & 0.0190 & 0.0190 & 0.2202 \\
3.9 & -7.1047 & -0.0773 & -0.1826 & -0.3324 & 0.0718 & 0.0718 & 0.0123 & 0.0123 & 0.2097 & 0.1851 & 0.0188 & 0.0188 & 0.2167 \\
4.0 & -7.1084 & -0.0814 & -0.1761 & -0.3304 & 0.0760 & 0.0760 & 0.0124 & 0.0124 & 0.2107 & 0.1835 & 0.0186 & 0.0186 & 0.2133 \\
4.1 & -7.1116 & -0.0850 & -0.1704 & -0.3286 & 0.0799 & 0.0799 & 0.0126 & 0.0126 & 0.2113 & 0.1820 & 0.0184 & 0.0184 & 0.2101 \\
4.2 & -7.1144 & -0.0880 & -0.1652 & -0.3269 & 0.0833 & 0.0833 & 0.0127 & 0.0127 & 0.2117 & 0.1805 & 0.0182 & 0.0182 & 0.2072 \\
4.3 & -7.1168 & -0.0906 & -0.1607 & -0.3255 & 0.0864 & 0.0864 & 0.0128 & 0.0128 & 0.2117 & 0.1791 & 0.0180 & 0.0180 & 0.2045 \\
4.4 & -7.1189 & -0.0927 & -0.1566 & -0.3242 & 0.0892 & 0.0892 & 0.0129 & 0.0129 & 0.2115 & 0.1778 & 0.0179 & 0.0179 & 0.2020 \\
4.5 & -7.1207 & -0.0944 & -0.1530 & -0.3231 & 0.0916 & 0.0916 & 0.0130 & 0.0130 & 0.2111 & 0.1766 & 0.0178 & 0.0178 & 0.1998 \\
4.6 & -7.1223 & -0.0958 & -0.1497 & -0.3221 & 0.0937 & 0.0937 & 0.0131 & 0.0131 & 0.2105 & 0.1753 & 0.0176 & 0.0176 & 0.1977 \\
4.7 & -7.1236 & -0.0968 & -0.1468 & -0.3213 & 0.0956 & 0.0956 & 0.0131 & 0.0131 & 0.2098 & 0.1742 & 0.0175 & 0.0175 & 0.1958 \\
4.8 & -7.1286 & -0.1167 & -0.1252 & -0.3164 & 0.1036 & 0.1036 & 0.0136 & 0.0136 & 0.2129 & 0.1735 & 0.0168 & 0.0168 & 0.1894 \\
4.9 & -7.0222 & -0.0264 & -0.2224 & -0.3165 & 0.0001 & 0.0001 & 0.0000 & 0.0000 & 0.1131 & 0.1051 & 0.0301 & 0.0301 & 0.2561 \\
5.0 & -7.0208 & -0.0252 & -0.3165 & -0.2237 & 0.0000 & 0.0000 & 0.0001 & 0.0001 & 0.1031 & 0.1118 & 0.0299 & 0.0299 & 0.2563 \\
\hline
\end{tabular}
\label{LiH_data}
\end{threeparttable}
\end{table*}

\end{document}